# WST, the wide-field spectroscopic telescope: progress on the design of the instruments

David Lee[a*], Joël D. R. Vernet[b], Elizabeth George[b], Omar Sqalli[b], Olaf Iwert[b], Alessandro Meoli[b], Vincenzo Mainieri[b], Philippe Dierickx[c], Alexandre Jeanneau[c], Corentin Cudennec[c], Paolo Franzetti[d], Olga Bellido[e], Roelof S. de Jong[e], Andreas Kelz[e], Aaron Omadutt[e], Steve Watson[a], Younes Chahid[a], Chris Waring[a], Laura Gibbs[a], Anmol Goyal[a], Dave Melotte[a], Lawrence Bissell[a], Jay Stephan[a], Oscar Gonzalaez[a], Sébastien Pernecker[f], Maxime Rombach[f], Malak Galal[f], Jonathan Wei[f], Jean-Paul Kneib[f], Joseph W. Barrow[g], Jon Lawrence[g], Will Saunders[g], Dimitri Buffat[h], Kjetil Dohlen[h], Andrea Tozzi[i], Anna Brucalassi[i], Simone D'Auria[i], Sofia Randich[i], Matteo Munari[j], Andrea Bianco[k], Christophe Yèche[l], Christophe Magneville[l], Etienne Burtin[l], Thomas Augusteijn[m], Jan Kragt[m], Ramón Navarro[m], Julia Bryant[n], Laurane Fréour[o] and Roland Bacon[c]

[a] UK Astronomy Technology Centre (United Kingdom),
[b] European Southern Observatory (Germany),
[c] Centre de Recherche Astrophysique de Lyon, Lyon (France),
[d] INAF-IASF Milano, Milan (Italy),
[e] Leibniz-Institut für Astrophysik Potsdam (Germany),
[f] Ecole Polytechnique Fédérale de Lausanne (Switzerland),
[g] Australian Astronomical Optics (AAO), Macquarie University, Macquarie Park, Sydney (Australia),
[h] Aix-Marseille Université, CNRS, CNES, LAM UMR7326, Marseille (France);
[i] INAF - Osservatorio Astrofisico di Arcetri, Florence (Italy),
[j] INAF - Osservatorio Astrofisico di Catania, Catania (Italy),
[k] INAF - Osservatorio Astronomico di Brera (Italy),
[l] University of Paris-Saclay, CEA-IRFU (France),
[m] NOVA optical infrared instrumentation group at ASTRON, Oude Hoogeveensedijk 4, Dwingeloo, (Netherlands),
[n] Sydney Institute for Astronomy, The University of Sydney (Australia),
[o] Department of Astrophysics, University of Vienna, Vienna (Austria).

## ABSTRACT

WST – the Wide-field Spectroscopic Telescope is a proposed new facility that will provide a transformational gain in spectroscopic survey capability over existing facilities. The WST is a 12 metre class telescope equipped with instrumentation to provide simultaneous observations in both multiple-object spectroscopy and integral field spectroscopy modes. This paper will describe the status of the instruments being designed for the WST.

The wide-field of view of the telescope, 2 degrees in diameter, will illuminate a fiber-fed multiple-object spectrograph capable of observing 30,000 objects at low spectral resolving power, and 2,000 objects at spectral resolving power of R = 40,000. An overview of the fiber positioner technology, and fiber relay system, will be provided. At the centre of the telescope's field of view an optical relay system will direct light to a large integral field spectrograph with a 3 arc-minute by 3 arc-minute field of view. An overview of the designs of the low-resolution, of the high-resolution, and of the integral field spectrographs will be presented. Technology advancements are being investigated to make the spectrographs more optically efficient and sustainable, such as CMOS detectors with low power cooling systems. The

[*] david.lee@stfc.ac.uk

WST facility will include an advanced calibration system capable of illuminating the spectrographs with a variety of light sources and providing metrology function for the fiber positioners. Finally, an overview of the overall layout of the instruments within the WST facility will be provided.

# 1. INTRODUCTION

The Wide-field Spectroscopic Telescope (WST) is an ambitious facility that will enable transformational science in many areas of modern astrophysics. The motivation and science drivers for WST are described elsewhere in the proceedings of this conference [1] and previously [2]. The WST will be a 12 metre class wide-field spectroscopic survey telescope with *simultaneous operation* of both Integral Field Spectroscopy (IFS) and Multiple-Object Spectroscopy (MOS) observing modes. In addition, the MOS capability will include both low- and high-resolution modes, known as MOS-LR and MOS-HR respectively. An early concept for a spectroscopic survey telescope is reported in [3].

The top-level requirements for WST's telescope and instrumentation are listed in Table 1. The initial set of instruments are specified to operate over the wavelength range 370 nm to 930 nm, but the telescope has been designed to operate over an extended wavelength range up to 1600 nm. This will enable the facility to be upgraded in future to include near-infrared instrumentation.

**Table 1. WST Top-Level Requirements**

| | |
|---|---|
| Telescope Aperture | 12 m (100 m$^2$) |
| Telescope Field of View | 2.0 degrees diameter, 3.1 degrees$^2$ |
| Telescope spectral range | 370 nm – 1600 nm |
| MOS-LR multiplex | 30,000 |
| MOS-LR spectral resolving power | 3,800 at 420 nm, 4,900 at 680 nm |
| MOS-LR spectral range | 370 nm – 930 nm, simultaneous |
| MOS-HR multiplex | 2,000 |
| MOS-HR spectral resolving power | 40,000 |
| MOS-HR spectral range | 370 nm – 930 nm (4 regions) |
| IFS field of view | 3 arc-minutes × 3 arc-minutes |
| IFS sampling | 0.25 arc-seconds × 0.25 arc-seconds |
| IFS spectral resolving power | 4,800 at 480 nm and 750 nm |
| IFS spectral range | 370 nm – 930 nm, simultaneous |
| IFS patrol field | 13 arc-minutes diameter |
| MOS and IFS simultaneous operation | |
| Target of Opportunity implemented at telescope and fiber level | |

The conceptual design of the WST building, telescope, and instrumentation is shown in Figure **1**. The telescope is a Cassegrain design, with a three-element field and atmospheric dispersion corrector providing a wide field of view of 2 degrees in diameter. Full details of the telescope optical design, including the atmospheric dispersion compensation, can be found in [4] which discusses early designs, and [5] which describes the design shown here. The ~F/3.3 MOS focal surface is located approximately 1.6 m behind the last element in the field corrector and is approximately 1.4 m in diameter. The MOS focal plate will be equipped with approximately 32,000 robotic fiber positioners. The low- and high-resolution MOS spectrographs will be located on, or below, the azimuth floor, whilst the IFS is in the base of the telescope building. Light is transmitted from the MOS focal surface to the spectrograph slit using optical fibers.

The field of view for the IFS is extracted from the center of the MOS focal surface by a fold mirror, shown in Figure 2, and is directed towards an optical relay which sends the IFS beam towards the Nasmyth platform. On the Nasmyth platform a set of mirrors relays an image 6 arc-minutes in diameter, inside of which is the 3 × 3 arc-minutes squared scientific field of view. This three-mirror relay propagates the light through the Nasmyth platform and refocuses it at the coudé focus. The Nasmyth mirror relay includes a deformable mirror to provide ground layer adaptive optics correction of the coudé image. Below the Nasmyth platform, the coudé section of the relay comprises fold mirrors, a field derotation system, and a field lens. The final coudé image, formed underneath the telescope in the area known as the "IFS station", is F/28.5. For more details refer to [5] and [6].

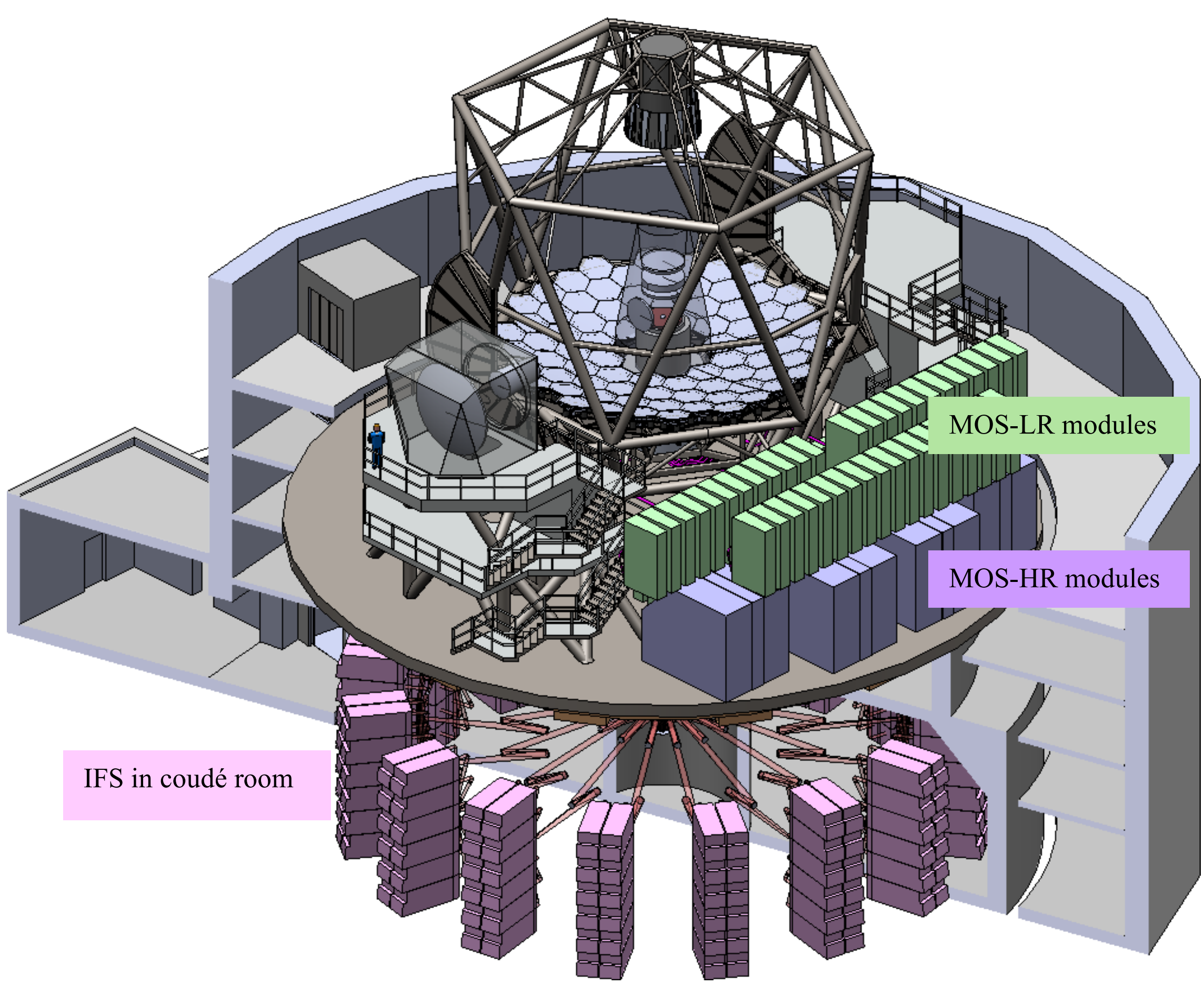

Figure 1. Conceptual layout of the WST facility showing the building, telescope, focal station, IFS relay, and IFS spectrograph modules located in the coudé room. In this concept the MOS-LR and MOS-HR spectrograph modules are located on the rotating azimuth floor. Image credit Gaston Gausachs.

WST will be equipped with a large flat-field screen mounted inside the dome. This will be illuminated with various calibration light sources to provide accurate and efficient calibration of the MOS and IFS instruments, as described in section 6.

Below the telescope azimuth floor is an area that could potentially house the MOS-LR and MOS-HR spectrographs in addition to services: electronics cabinets, cables, cooling pipes, etc. This level also accommodates some of the Coudé focus relay mirrors which must co-rotate with the telescope. The IFS field derotation system is housed within the central pier of the telescope. The layout of the building will be optimized for handling and maintenance given the size and complexity of the instrumentation suite. Primary mirror segments will be removed and replaced by a deployable crane. Other cranes and lifts will be necessary for positioning of the IFS and MOS modules.

Detailed image quality and throughput information is gathered for each sub-system: telescope, fibers, instruments, and detectors. This information is collated in a spreadsheet to enable calculation of the final end-to-end system throughput. WST system throughput budget MOS-LR and MOS-HR end-to-end system throughput is expected to be 35% and 18% respectively at wavelength 650 nm. IFS system throughput is predicted to peak at ~40% at 800 nm. This information is passed to the exposure time calculator [7] where it is used in the calculation of limiting magnitude sensitivity for WST.

## 2. FIBER POSITIONERS AND FRONT END

The design of the fiber positioning system for WST is being developed by four collaborating organizations: Australian Astronomical Optics (AAO), Ecole Polytechnique Federale de Lausanne (EPFL), Leibniz Institute for Astrophysics Potsdam (AIP), and United Kingdom Astronomy Technology Centre (UKATC).

Four different fiber positioner conceptual designs were initially developed for the WST MOS, as described in [8]. The first design, proposed by AIP is called FLEX and consists of a flexible arm, with three motors, providing positioning in three degrees of freedom, x, y, and focus. More information can be found in [9], [10] and [11]. A second design, proposed by EPFL, uses mature Phi-Theta robotic positioner technology, with two rotating axes, and is also described in [8]. The use of Phi-Theta fiber positioners is well established and already in use in instruments such as DESI [12] and PFS [13]. A range of Phi-Theta actuators has been developed by EPFL and test results for these fiber positioners are reported in [14]. AAO proposed a third option to use tilting spine technology, with two degrees of freedom of motion, as already successfully used on 4MOST. A fourth design, developed by UKATC, uses R-theta motion, with one radial motor and one rotating motor. This is the only option to include built encoders for position feedback [15]. It is interesting to note that R-theta fiber positioning was also originally considered for 4MOST [16].

Following a period of fiber positioner technology development the baseline technology to be adopted by the WST was selected using a trade study process. FLEX positioners were selected as the baseline due to their large patrol field (>30 mm) whilst having small pitch (~7 mm). More mature technology of tilting spines and theta-phi positioners will continue to be developed as a backup solution.

There are two conceptual designs for the layout of the front end which locates the fiber positioners on the telescope's focal surface, as shown in Figure 2. The front end conceptual design on the left of Figure 2 uses inline modules with each module containing hundreds of fiber positioners. This type of focal plane layout has already been successfully used with tilting spines. The front end conceptual design on the right of Figure 2 uses modular triangular rafts with each raft containing 63 positioners. This concept is based on one already developed for MegaMapper [17] with designs for WST reported in [18].

Both the inline modules and triangular rafts can be arranged, with high fill-factor, to form a large hexagon that fills the field of view. The remaining field of view, within the 2 degrees diameter but outside the hexagon, is used to accommodate acquisition and guiding, curvature wavefront sensors, and Shack-Hartmann wavefront sensors. In Figure 2 the entrance aperture of acquisition and guiding and curvature wavefront sensors are indicated by red squares. The three blue rectangles indicate the entrance aperture of the Shack-Hartmann wavefront sensors. The concept with triangular modules also shows the triangular electronics modules below the focal surface. The cylinder in the center of the focal surface indicates the volume envelope of the M3 IFS pick-off mirror.

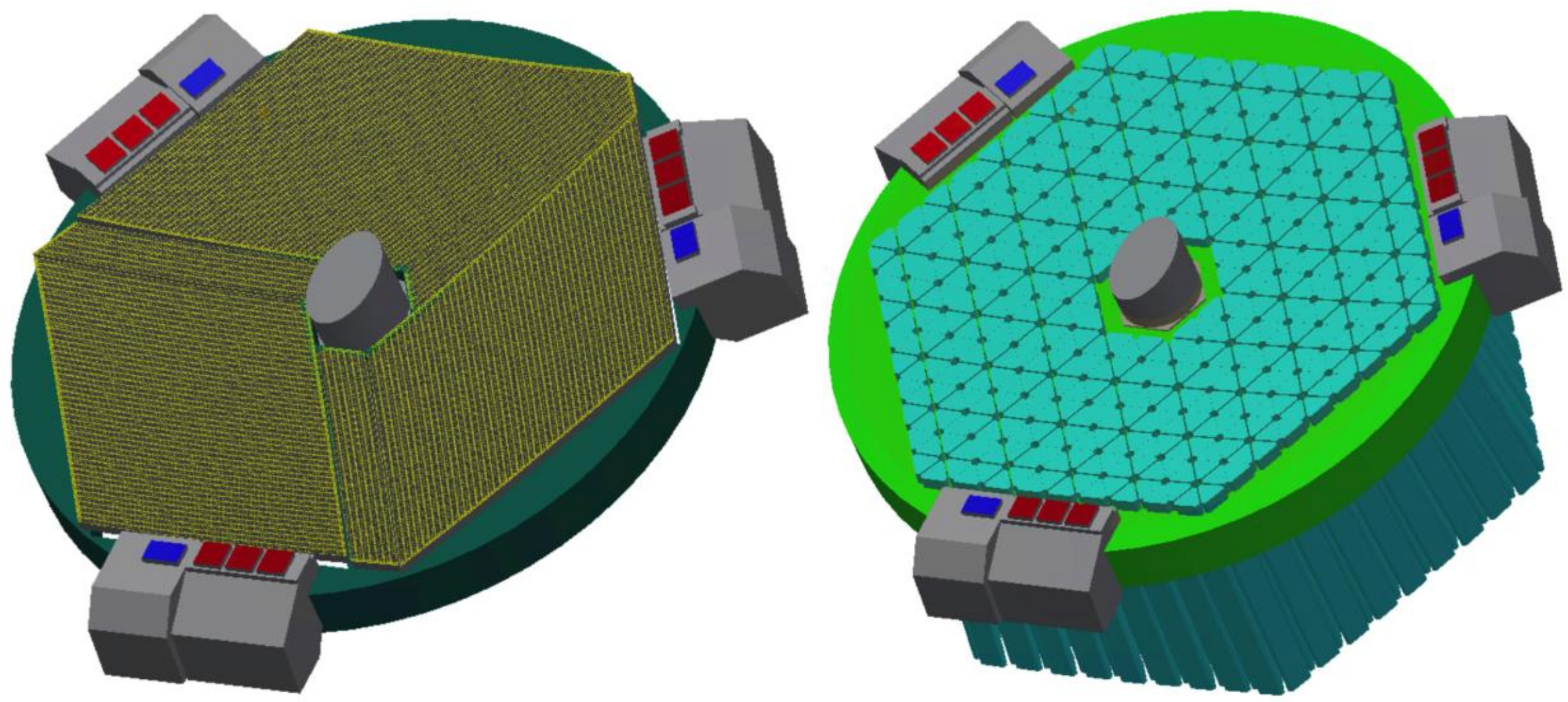

Figure 2. Two conceptual designs for the WST focal plane layout. The design on the left uses inline modules to tile the focal surface whilst the design on the right uses triangular rafts. Image credit Younes Chahid.

During target acquisition the position of each fiber will be measured using a metrology camera located on the telescope structure in a position to view the MOS focal surface. Full coverage of the entire MOS field of view will likely need multiple metrology cameras. During metrology measurements the fiber slit will be back illuminated by a light source located within the MOS spectrograph. The wavelength of the metrology source will be chosen so as not to be visible by the IFS during simultaneous observations.

## 3. FIBERS

The WST fiber layout will be split into two or three lengths of fiber as follows. The first length is the positioner fiber that takes the light from the WST MOS focal surface to the first connector, located at the back of the front end. Both MOS-LR and MOS-HR positioner fibers are the same, with an on sky aperture of 1.0 arc-seconds, as this simplifies the design of the positioner as only one type of fiber needs to be accommodated. The positioner fibers terminate at a connector, at the interface between the front end structure and the telescope structure. A connector at the front end interface is considered critical to integration and ongoing maintenance of the front end sub-system.

Modified off-the-shelf fiber connectors are already successfully used in several existing MOS instruments, e.g. 4MOST, and provide throughput of ~90%. This has been identified as an area where performance improvements can be made, and the development of high-performance fiber connectors is included in the WST technology development plan.

Following the first connector there will be a fiber bundle of significant length to transport the light through the telescope structure and potentially through two cables wraps. Several alternative fiber routes from front end to spectrograph are being explored to minimize the fiber length and hence maintain good throughput in the blue. A second optional fiber connector will be located adjacent to the MOS spectrograph and would permit a break between the long second section of fiber and a third shorter section into the spectrograph. The fibers terminate at the MOS slit, which will be composed of an array of slitlet blocks each containing a group of fibers.

For the MOS-HR to achieve high spectral resolving power requires the use of fiber splitting so that the spectrograph is illuminated with an appropriately sized input fiber. Fiber splitting, from one large (200 μm core diameter) to seven smaller (77 μm core diameter) fibers, will be done using a photonic lantern to provide good mode matching and hence high efficiency. The development of high throughput photonic lanterns, with low focal ratio degradation, is included in the WST technology development plan.

## 4. MULTIPLE-OBJECT SPECTROGRAPH

WST will have two types of multiple-object spectrographs to provide low- and high-resolution observations: MOS-LR and MOS-HR respectively. Several design options for the MOS-LR have been investigated and are reported in [19] and [20] and the MOS-HR optics and mechanical design options are reported in [21], [22] and [23].

A conceptual design for the MOS-LR spectrograph is shown in Figure 3. The optical configuration consists of a curved slit with coupled lens, off-axis reflective collimator, transmissive disperser, five element refractive camera, and a flat detector. The design shown has four wavelength channels the bandpass of which is defined by three dichroic beam-splitter filters.

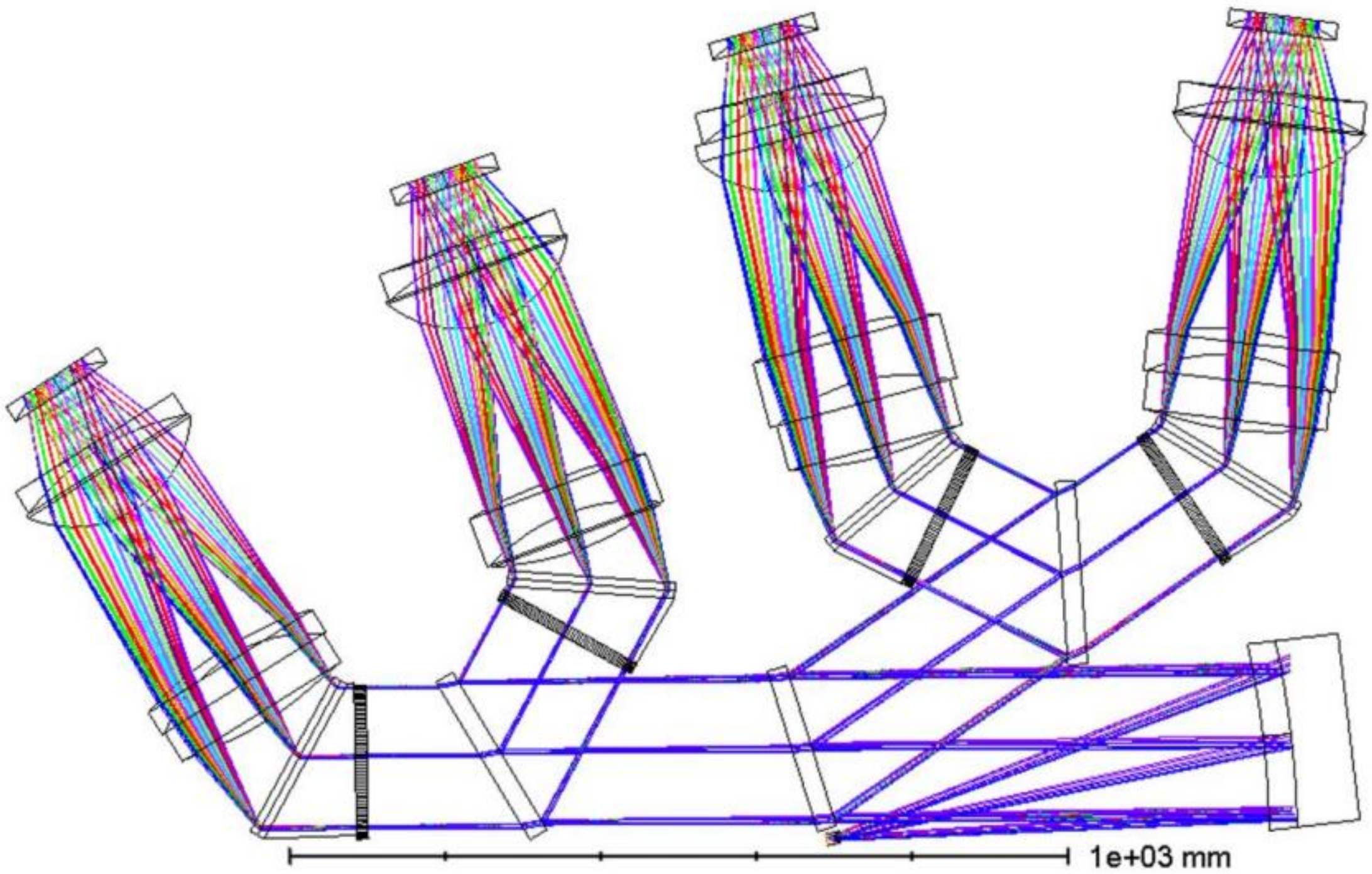

Figure 3. Optical design concept for a four channel MOS-LR low resolution multiple-object spectrograph. The position of the entrance slit is at the base of the figure.

A preliminary conceptual optical design for the MOS-HR spectrograph is shown in Figure 4. The optical configuration consists of a curved slit, on-axis collimator mirror with refractive field corrector, transmissive disperser, catadioptric cameras, and a two by two detector mosaic with 6k × 6k detectors with 10 µm pixels. Each MOS-HR spectrograph can observe four pass bands, and 8 spectrograph modules are needed to provide the required multiplex of 2,000. More information regarding the MOS-HR can be found in references [21], [22], and [23]. The location of the MOS-HR modules within the telescope building is under careful consideration. To keep the fiber length to a minimum the MOS-HR modules may be located next to the telescope on the azimuth floor, as illustrated in Figure **1**. From the point of view of thermal stability, it might be preferable to locate the MOS-HR modules below the telescope azimuth floor, but this would require longer fibers.

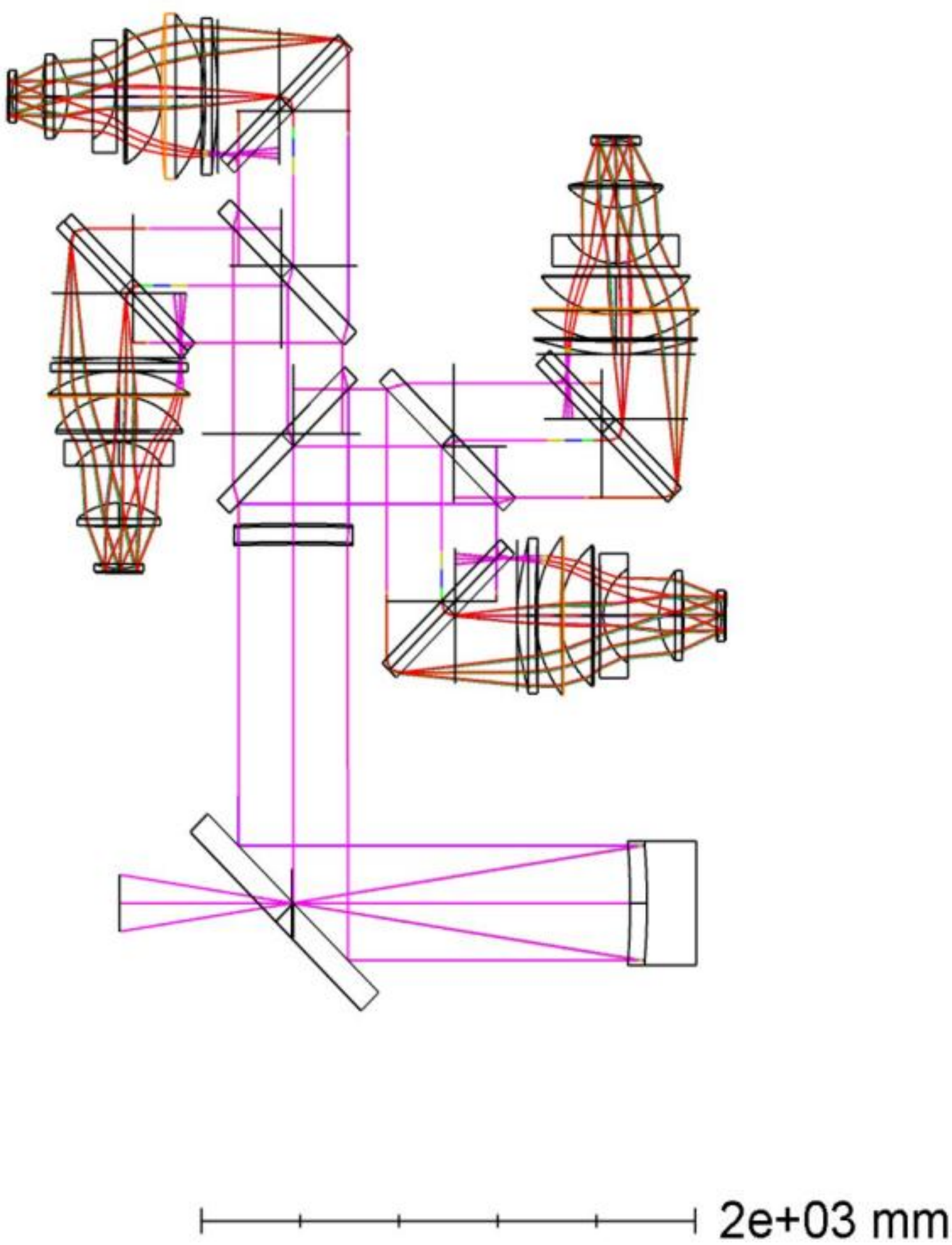

Figure 4. Optical design concept for the WST MOS-HR high-resolution spectrograph. The spectrograph entrance slit is mounted in a slot in the back of the fold mirror. The folded collimator configuration minimizes the overall volume.

## 5. INTEGRAL FIELD SPECTROGRAPH

The WST integral field spectrograph will be located in the base of the WST building, in the stationery coudé room, as shown in Figure **1** and Figure 5. The final few elements of the IFS relay optics are shown in Figure 5, a fold mirror, the K-mirror to provide image derotation, and a field lens immediately prior to the IFS input focal surface.

A first stage field splitter is mounted at the IFS focus and this splits the field of view into sixteen sub-fields. These sixteen beams are then reimaged onto the second stage field splitter by the first stage relay optics, located in the center of the IFS station. The second stage field splitter further divides the field of view into twelve sub-fields, each of which is then reimaged into the IFS spectrograph module.

Sixteen blocks, each containing twelve IFS modules, are arranged symmetrically around the outer enclosure of the IFS station. The overall layout of the IFS is based on the MUSE instrument in operation on the European Southern Observatory's Very Large Telescope [24]. The electronics and cooling systems for the one hundred and ninety-two integral field spectrographs will be extremely complex and require significant volume. Technology development studies are underway to investigate options to miniaturize and simplify these areas, as mentioned in section 7.

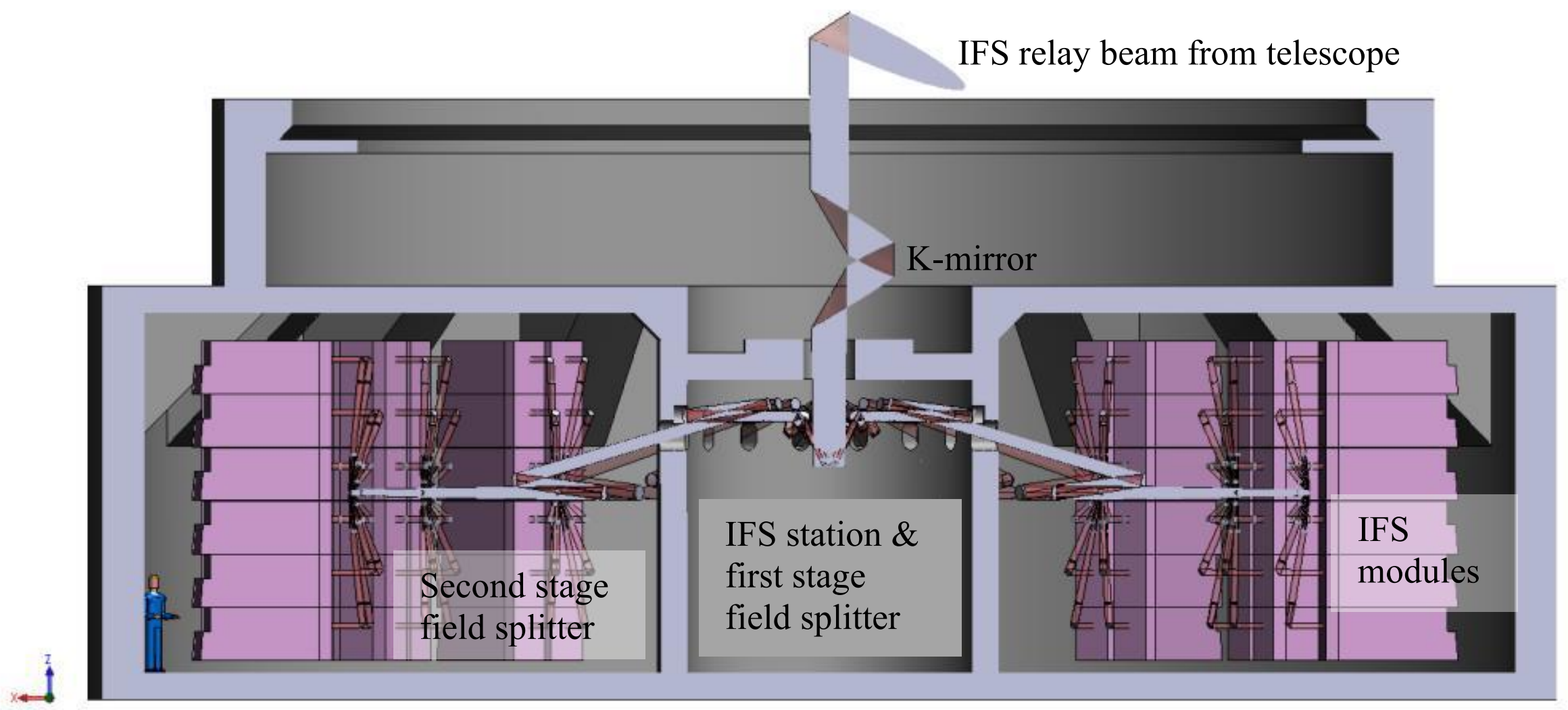

Figure 5. Conceptual CAD rendering of the IFS optical path through the K-mirror to the IFS two stage field splitter and to the IFS modules (coloured pink). Image credit Gaston Gausachs.

Figure 6 shows a conceptual optical design of an integral field spectrograph with two wavelength channels. The wavelength split is achieved with a dichroic beam-splitter filter. The optical design consists of a reflective collimator with corrector lenses, a transmissive disperser, and a four-element camera. The optical design assumes the use of a curved detector, which reduces the number of lenses needed in the camera, whilst still achieving the image quality requirements. Full details of the design of the IFS can be found in [25] and [26]. A significant challenge for the design of the IFS will be the large-scale industrial manufacturing of the many IFS optical modules.

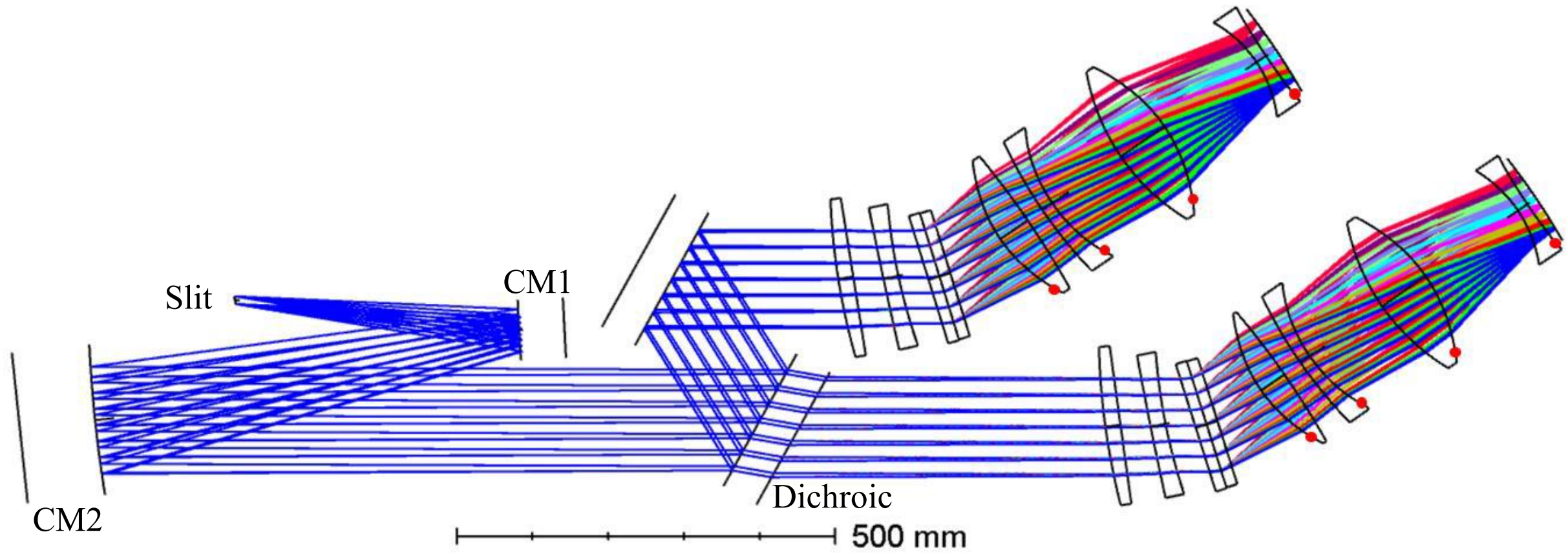

Figure 6. Optical ray-trace diagram of a proposed IFS spectrograph with a curved detector. CM1 and CM2 indicate the two mirrors that form the collimator. A small red circle indicates an aspheric surface. Each pink block shown in Figure 6 contains one of these IFS modules.

## 6. CALIBRATION

The ability to efficiently and accurately calibrate WST instruments is being included in the baseline design of the facility. Efficient calibration implies a minimum time overhead and the requirement to avoid night time calibrations if necessary. Accurate calibration implies the performance of each instrument should be well characterized and stable.

A large dome screen will be used to illuminate the entire telescope, focal surfaces, and all instruments simultaneously. Other methods of illumination of the telescope were considered but the dome screen has the key advantage that it correctly illuminates the entire telescope entrance pupil and hence the calibration images measured by the instruments will closely mimic on-sky measurements. Other methods of illumination, that do not correctly illuminate the pupil, cause the calibration images to have small differences compared to those measured on sky. Twilight sky exposures will also be used to measure instrument performance as these provide ideal illumination and can be measured at a variety of zenith angles.

A disadvantage of using a large dome screen is that performing calibration requires the telescope to be pointed at the screen which is a time-consuming overhead during night-time observations. To enable fast night-time observations the telescope spiders will be equipped with light sources or “light-stripes”, called light sabers on 4MOST [27], that can be used as any telescope position, night or day.

The dome screen will be illuminated with quartz tungsten halogen lamps, to provide flat-field illumination, and gas discharge lamps, e.g. Argon, to provide wavelength calibration. The density of spectral features available from typical gas discharge lamps may not be sufficient to provide wavelength calibration for the MOS-HR and hence hollow cathode lamps will also be needed. Tunable laser light sources will be used to provide monochromatic spectral lines for line spread function (LSF) and fiber spread function (FSF) calibration.

All MOS spectrographs will be fitted with simultaneous calibration fibers, approximately nine per slit, and illuminated with a combination of gas lamps: hollow cathode lamps as the absolute reference, and a white light source illuminating a Fabry-Perot interferometer to provide a regular grid of calibration lines.

MOS and IFS spectrographs will include three different colored light sources that directly illuminate the detector to provide an approximately uniform detector flat-field.

The IFS will have an independent calibration system, providing wavelength calibration and flat-field illumination, with light entering the IFS via a deployable screen, or mirror, inserted into the IFS relay. A deployable screen, or mirror, also serves a dual purpose as an IFS shutter, blocking the entire field of view. An additional calibration requirement for the IFS is geometric calibration, which will be achieved by projecting an image of a grid target into the IFS, via a fold mirror inserted into the IFS relay beam [28].

## 7. TECHNOLOGY DEVELOPMENT

To maximise the performance of the WST a forward look has been taken to identify future technology trends and advancements that can be used in the WST instruments. Several enabling technologies have been identified, and these will be included in the WST technology development plan or roadmap. These key areas are:

1. Optical coatings, described in [29].
2. CMOS detectors, described in [30].
3. Disperser technology, described in [31]and [32].
4. Photonic lanterns as fiber splitters.
5. High throughput multi-mode fiber connectors.
6. High efficiency cooling systems [33].
7. Calibration illumination systems.

Following an extensive trade study exercise the baseline detector technology selected for WST is CMOS as it offers several potential performance advantages over CCDs. The advantages of CMOS include low read noise, less time to read out, and the possibility of new observation modes enabled by non-destructive readout. The required detector format for both the MOS-LR and IFS has converged on the same specification: 6k × 6k with 15 µm pixels. The IFS has two spectrograph designs that use either flat or curved detectors. More information on detector development for WST can be found in [30].

All the WST's spectrographs are designed to use transmissive dispersers. The collimated beam size in the IFS and MOS-LR is within existing capability of several grating manufacturing companies and presents no major technical challenge. The collimated beam size in the MOS-HR is more challenging, but it is still within the manufacturing capability of two companies. In all cases peak diffraction efficiency above 90% is theoretically predicted. Both volume phase holographic and etched lithographic grating technology are under consideration, and full details can be found in [31] and [32].

The connectorization of many thousands of optical fibers for the WST is another technical challenge. High efficiency, low focal ratio degradation, fiber connectors will be needed that potentially make hundreds of fiber-to-fiber connections in a single device. In addition to the connectors the MOS-HR fibers will also need photonic lanterns to provide fiber splitting. Development of optical devices to meet these requirements is the subject of on-going research activity within the WST consortium partners.

Sustainability has been an important consideration for the design and operation of WST and has been included as a factor in the choice of detector [33]. The ability of CMOS devices to operate at higher temperatures than CCDs means that an energy efficient Carbon Dioxide gas and Krypton gas cooling system can be used instead of less efficient linear pulse tube or liquid nitrogen cooling systems.

## 8. SUMMARY & CONCLUSIONS

This paper provides an overview of the Wide-Field Spectroscopic Telescope and the conceptual design for the instrumentation. Many of the conceptual designs presented are based on the heritage of existing instruments, such as MUSE and 4MOST, and are considered technically feasible. As a future facility, the conceptual design of the WST instruments will be able to take advantage of rapidly developing new technology, such as optics and electronics, to offer performance advantages over existing facilities.

More information can be found at the WST website: https://wstelescope.eu/

## ACKNOWLEDGEMENTS

This project has been funded by the European Union's Horizon Europe research and innovation programme under grant agreement No. 101183153.